\documentclass[]{spie}  

\usepackage{amsmath,amsfonts,amssymb}
\usepackage{graphicx}
\usepackage{epstopdf}
\usepackage[colorlinks=true, allcolors=blue]{hyperref}
\usepackage[tight,normalsize,sf,SF]{subfigure}

\title{Multimodal Brain Visualization}

\author[]{Saad Nadeem}
\author[]{Arie Kaufman}
\affil[]{Department of Computer Science, Stony Brook University, Stony Brook, NY, 11794, USA}

\authorinfo{Preprint for accepted SPIE Medical Imaging Conference 2016 manuscript. DOI: 10.1117/12.2217003}

\begin{document}
\maketitle

\begin{abstract}
Current connectivity diagrams of human brain image data are either overly complex or overly simplistic. In this work we introduce simple yet accurate interactive visual representations of multiple brain image structures and the connectivity among them. We map cortical surfaces extracted from human brain magnetic resonance imaging (MRI) data onto 2D surfaces that preserve shape (angle), extent (area), and spatial (neighborhood) information for 2D (circular disk) and 3D (spherical) mapping, split these surfaces into separate patches, and cluster functional and diffusion tractography MRI connections between pairs of these patches. The resulting visualizations are easier to compute on and more visually intuitive to interact with than the original data, and facilitate simultaneous exploration of multiple data sets, modalities, and statistical maps.
\end{abstract}


\section{INTRODUCTION}
\label{sec:intro}  
The human brain is notoriously complex in its shape and connectivity, even at scales at which magnetic resonance imaging (MRI) can measure. Visualization is a natural means of exploring possible relationships across scan types (multimodal data) or statistical variables (multivariate data). However, visualizing brain image data in the original form of the brain can result in dense, cluttered, and uninterpretable images, whereas visualizing an abstract representation of information extracted from the same data can result in over-simplified images that are too removed from their source to be intuitive and meaningful \cite{margulies:2013}. In this article, we introduce novel methods for representing brain image data that are conducive to interactive visualizations that retain some of the benefits of the data in their original space while taking advantage of the simplicity and elegance of more abstract formats. To attain this balance, we map cortical surfaces extracted from human brain image data onto 2D surfaces that preserve shape (angle), extent (area) and spatial (neighborhood) information, split these surfaces into separate patches, and cluster connections between pairs of these patches. We maintain that the resulting visualizations are more intuitive and comprehensible than conventional native brain images or information graphics such as abstract network diagrams. The reason we are interested in visualizations of brain image data that are intermediate between realistic and abstract representations is not only that they could be easier to compute on than the original data and more visually intuitive to interact with than the abstract form, but they also have the potential to make it easier to simultaneously explore multiple data sets, modalities, and statistical maps.

\section{MULTIMODAL BRAIN VISUALIZATIONS}
Our focus in this work is to open up avenues to study the brain in simpler topological domains (disk or sphere), keeping the integrity of the original data. This will allow the user to explore the brain in the native and transformed space seamlessly, and hence be able to confirm the findings from the transformed space in the native space and vice versa. Apart from providing an easier way to visualize the brain data, the transformation to simpler topological domains also allows easier ways to compute on the brain data. In the following sections, we outline the details for our pipeline and the resultant visualizations.

\subsection{Pipeline}
\label{sec:pipeline}
\begin{figure}
\centering
\includegraphics[width=0.9\linewidth]{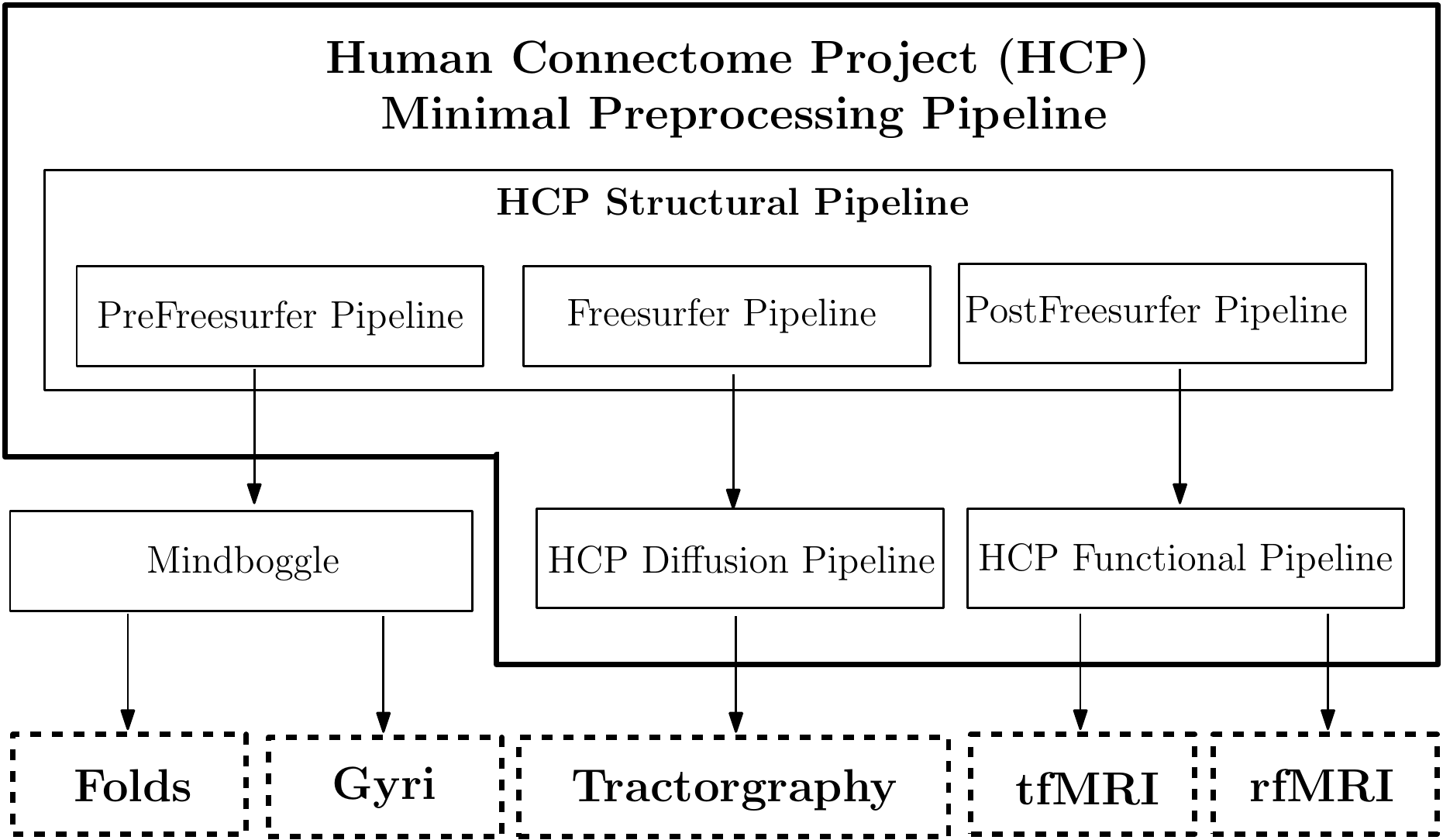}
\caption{\emph{Multimodal Brain Data Pipeline}. HCP MRI datasets are run through the Freesurfer \cite{Freesurfer:2012} pipeline to produce surfaces and volumes extracted from the initial MRI data. This raw data is then run through ANTS \cite{ANTS} and Mindboggle \cite{Mindboggle} to produce cortical features (folds and gyri). Simultaneously, the HCP Minimally Processed pipeline \cite{HCPminimal:2013} outputs pre-processed myelin maps, 68 tasks based fMRI analysis, functional connectivity map with full correlation and after regression of the mean gray timecourse from rfMRI data. The pre-processed Diffusion MRI data from the HCP Minimally Processed pipeline is used to recover fiber tracts from DSI Studio \cite{DSIStudio}, using deterministic tractography.}
\label{fig:pipeline}
\end{figure}

\begin{figure}[h]
\begin{center}
\begin{tabular}{cccc}
\includegraphics[width=0.27\linewidth]{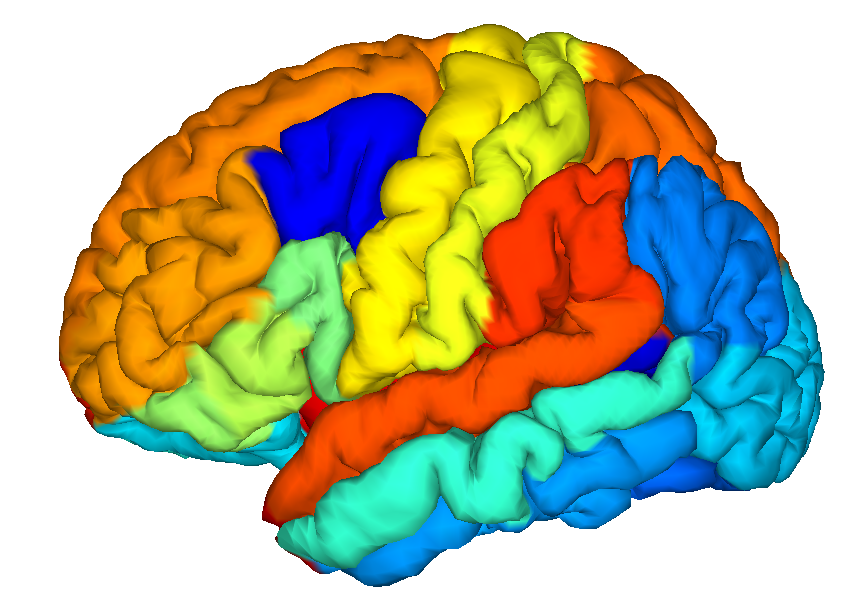}&
\includegraphics[width=0.26\linewidth]{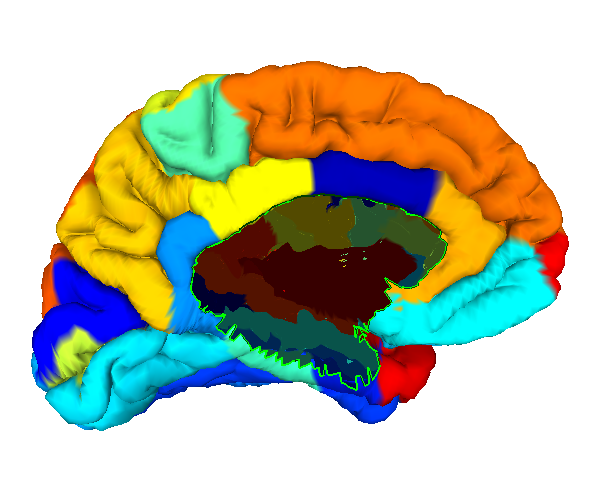}&
\includegraphics[width=0.215\linewidth]{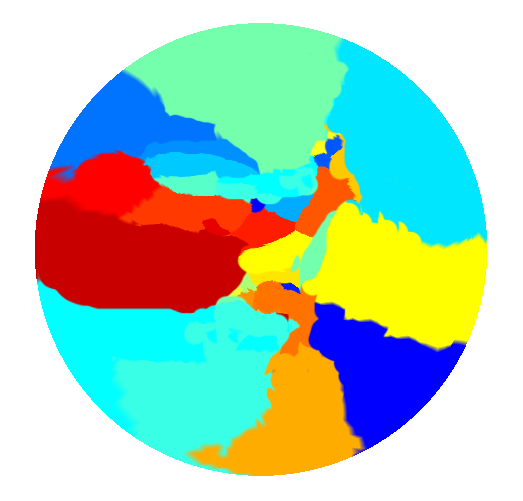}&
\includegraphics[width=0.20\linewidth]{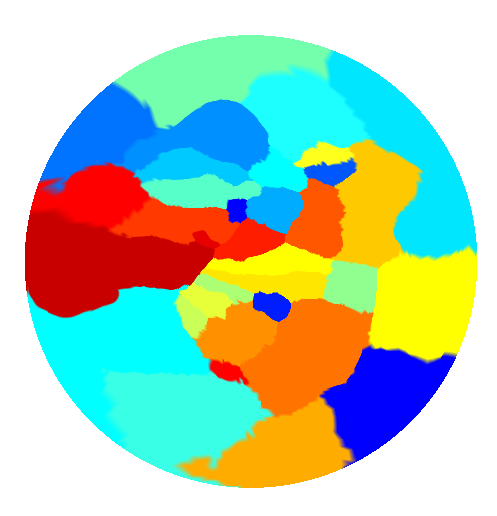}\\
(a) & (b) & (c) & (d)\\
\end{tabular}
\end{center}
\caption{\emph{Angle and area preservation comparison}. (a) Superior view of the left brain hemisphere. (b) Medial view of the left brain hemisphere with the colormap showing different gyri and the removed region (highlighted with green boundary) for the mapping in (c) and (d). (c) The angle-preserved mapping of (a) \& (b) with the vertices on the green boundary in (b) representing the boundary vertices. (d) The mapping of (a) \& (b) with angle preservation and area correction. \emph{Note: The colormap for the gyri is arbitrary and is only there to distinguish between different gyri. This will be the convention used in all figures to distinguish individual cortical surface feature regions.}}
\label{fig:angle_area_comparison_brain}
\end{figure}

We use the Human Connectome Project (HCP) \cite{HCPoverview:2013} data and a minimal preprocessing pipeline \cite{HCPminimal:2013} along with Mindboggle \cite{Mindboggle}, ANTS \cite{ANTS}, and DSI Studio \cite{DSIStudio} to obtain the highest fidelity structural, functional, and diffusion data for our visualizations. The transformation of the brain data from the initial brain MRI scans to the registered brain cortical surfaces (triangular meshes), tractography data, and functional clusters is outlined in Figure \ref{fig:pipeline}. The HCP pipeline produces separate left and right brain cortical surfaces to allow their exploration in isolation. We remove the extraneous region added to these resultant cortical surfaces (Figure \ref{fig:angle_area_comparison_brain}b) to allow for easier disk and spherical mapping, as demonstrated in the next section.

\setlength{\tabcolsep}{0.6pt}

\begin{figure}[h]
\begin{center}
\begin{tabular}{ccc}
\includegraphics[width=0.25\linewidth]{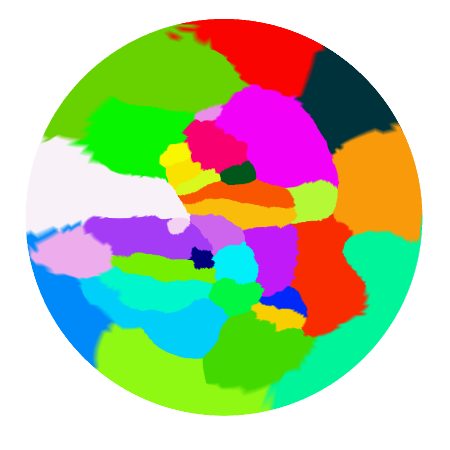}&
\includegraphics[width=0.25\linewidth]{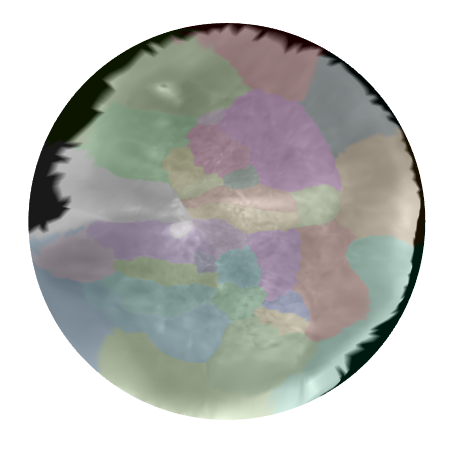}&
\includegraphics[width=0.1\linewidth]{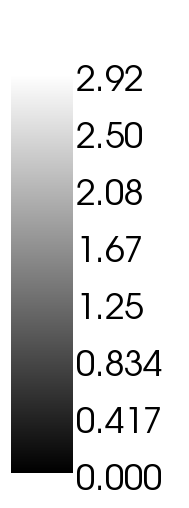}\\
(a) & (b) & \\
\end{tabular}
\end{center}
\caption{\emph{Myelin Map}. (a) The angle preservation and area correction map of a left brain hemisphere with gyri, represented with an arbitrary color map. (b) Myelin map overlaid on (a).}
\label{fig:myelin_map}
\end{figure}

\subsection{Disk and Spherical Mapping}
We use 2D angle-area preserving mapping \cite{XinZhao:2013} to transform the complex 3D geometry of a right or left brain hemisphere (computed from our pipeline \ref{fig:pipeline}) into a simpler 2D disk map, as shown in Figure \ref{fig:angle_area_comparison_brain}. The simultaneous angle and area preservation of the brain hemispheres is critical to study the overlaid statistical maps from the functional MRI data since these are of low resolution and the smallest distortions in area or angle can lose critical brain activity information. Moreover, the angle-area preserving mapping retains the original brain geometry information as closely as possible, hence allowing it to be mapped back to the original 3D geometry using simple sampling techniques. We demonstrate the 2D mapping overlaid with myelin deposits, which has been associated with autism and Alzheimer's, in a grayscale colormap in Figure \ref{fig:myelin_map}.

\begin{figure}[h]
\begin{center}
\includegraphics[width=0.55\linewidth]{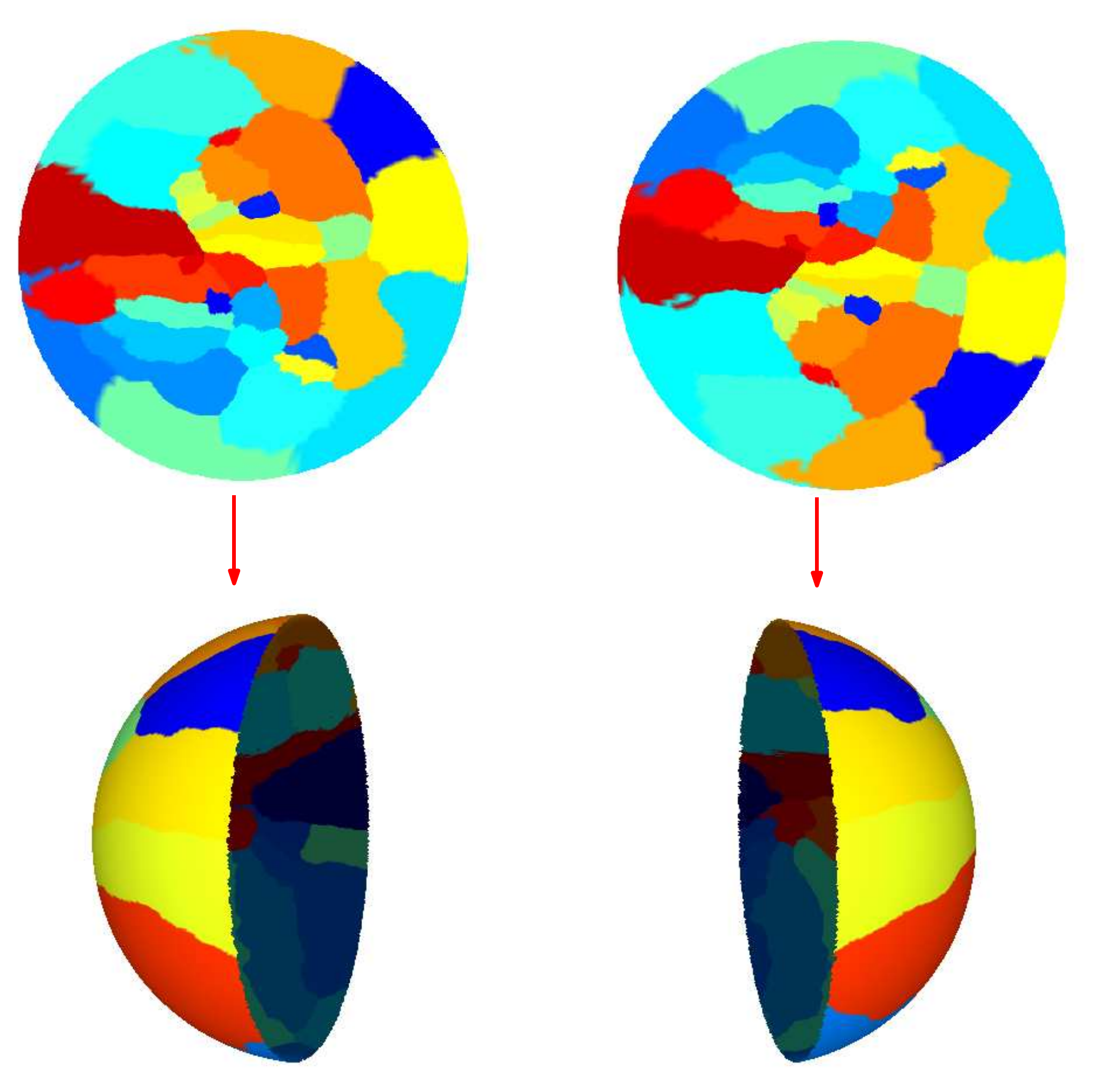}
\end{center}
\caption{\emph{Spherical Map}. The area-angle preservation disk maps of left and right brain hemisphere from Figure \ref{fig:angle_area_comparison_brain} are stereographically projected and aligned along the boundary vertices to create a single spherical map.}
\label{fig:spherical_map}
\end{figure}

We also present a spherical mapping approach to combine the two brain hemispheres from our pipeline using inverse stereographic projection and alignment along the boundary vertices, as shown in Figure \ref{fig:spherical_map}. The spherical mapping allows us to visualize the complete brain -- combined left and right hemispheres -- (which is a genus-0 surface) in its true parametric domain, that is, a sphere. We can now add diffusion, functional and other information onto these simpler topological domains, as demonstrated in the next section.

\begin{figure}[h]
\begin{center}
\begin{tabular}{ccc}
\includegraphics[width=0.25\linewidth]{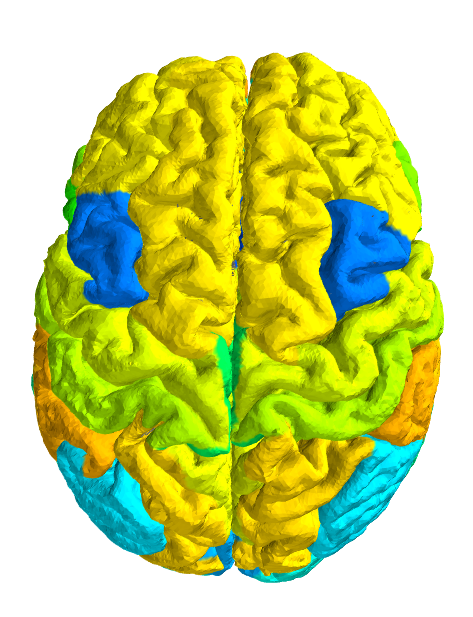}&
\includegraphics[width=0.25\linewidth]{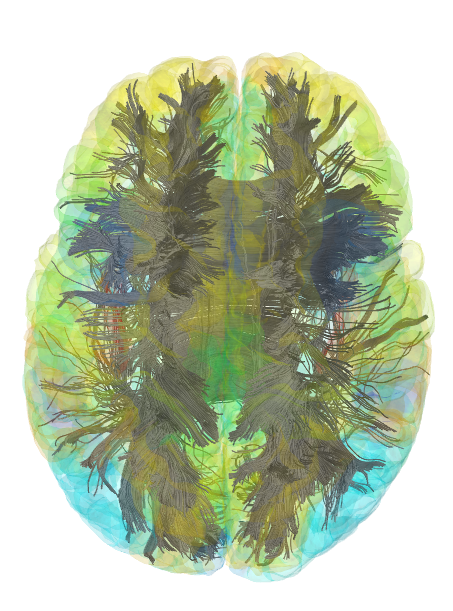}&
\includegraphics[width=0.25\linewidth]{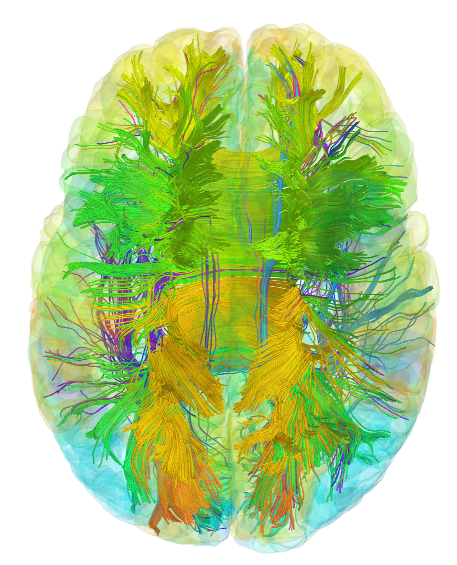}\\
(a) & (b) & (c)\\
\end{tabular}
\end{center}
\caption{\emph{QuickBundles \cite{garyfallidis:2012} clustering visualization}. (a) The original surface mesh. (b) Tractography information added to (a) using the diffusion MRI data from HCP. (c) Clustering information added to (b) where different colors represent different clusters.}
\label{fig:qb_clustering}
\end{figure}

\begin{figure}[h]
\begin{center}
\begin{tabular}{cc}
\includegraphics[width=0.35\linewidth]{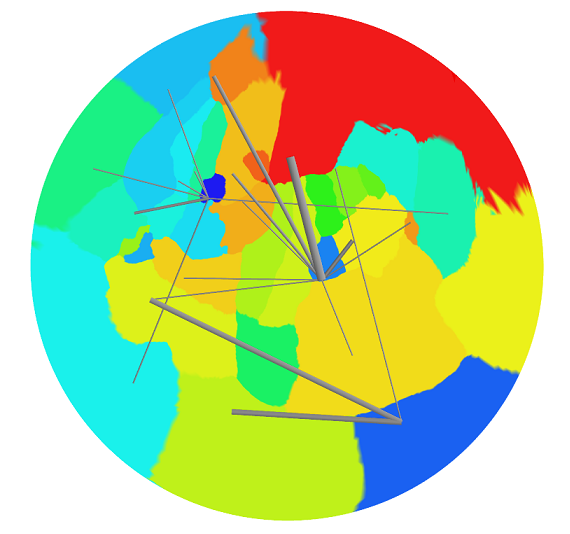}&
\includegraphics[width=0.5\linewidth]{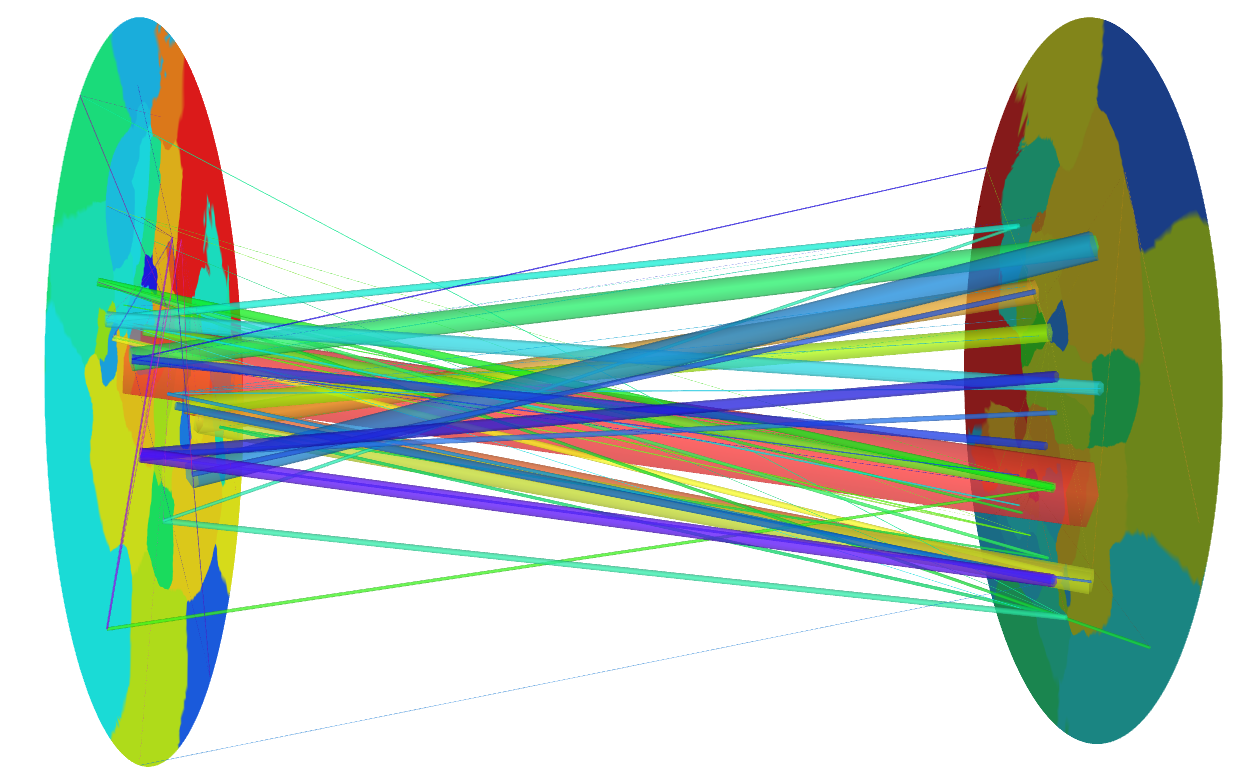}\\
(a) & (b)\\
\end{tabular}
\end{center}
\caption{\emph{Anatomical connectivity diagram for the entire brain with surface features (gyri) overlaid onto circular discs.}
The angle preservation and area correction mapping of left and right brain hemispheres on discs and the corresponding connections, within
a single hemisphere (a) and between the two hemispheres (b), computed based on the coalescing of individual fiber tract clusters from Figure \ref{fig:qb_clustering}.}
\label{fig:anatomical_connectivity}
\end{figure}

\begin{figure}
\centering
\includegraphics[width=1\linewidth]{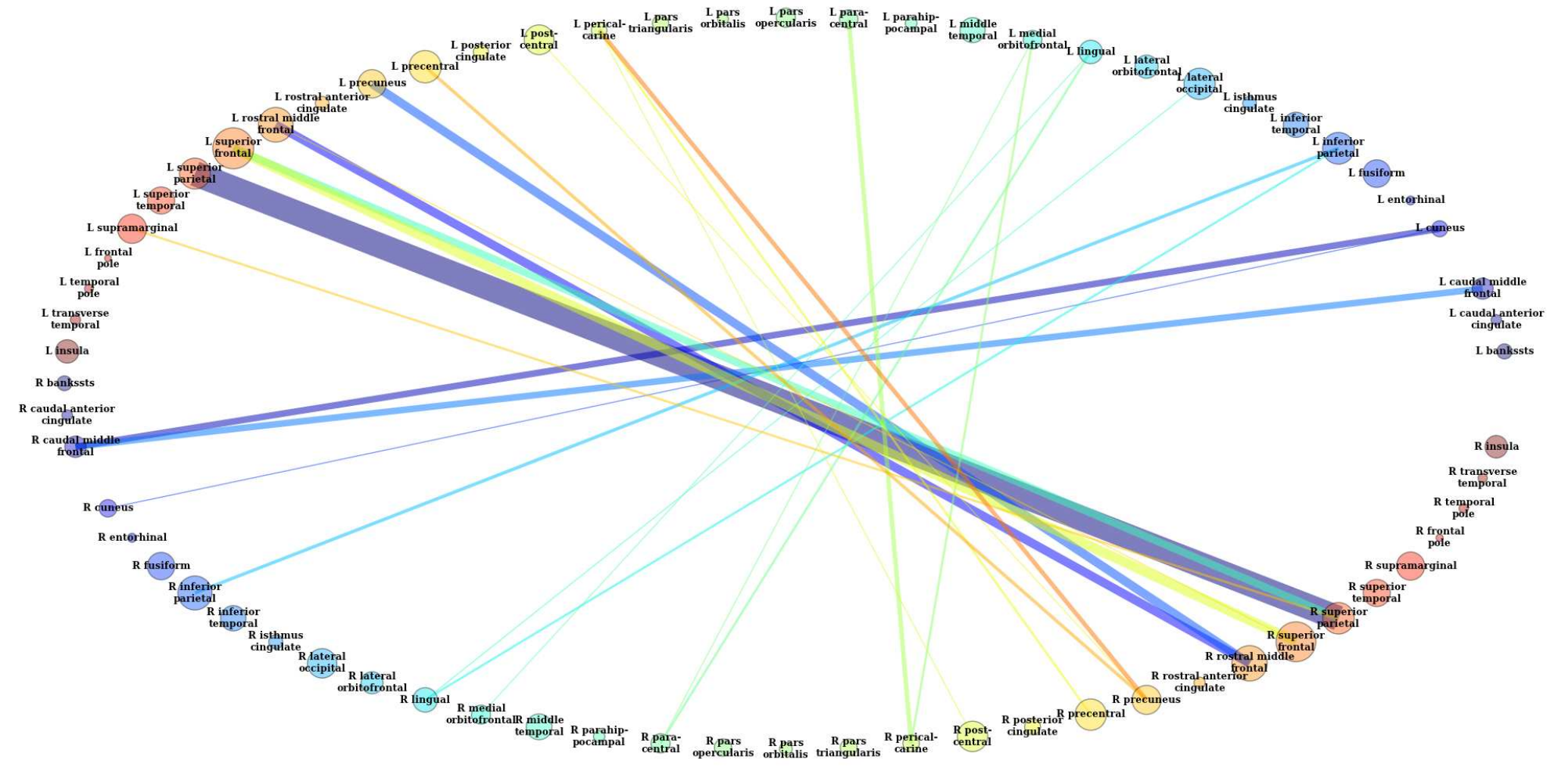}
\caption{\emph{Network diagram for the anatomical connectivity with surface features (gyri) color-encoded with the same colormap as in Figure \ref{fig:anatomical_connectivity}}. This visualization shows the same information in Figure \ref{fig:anatomical_connectivity}, but with only relative area information and connectivity information. \emph{Note: A different colormap for connections is used compared to Figure \ref{fig:anatomical_connectivity} to make the connections more distinguishable in 2D}.}
\label{fig:complete_brain_conn_network}
\end{figure}

\begin{figure}
\begin{center}
\begin{tabular}{ccc}
\includegraphics[width=0.3\linewidth]{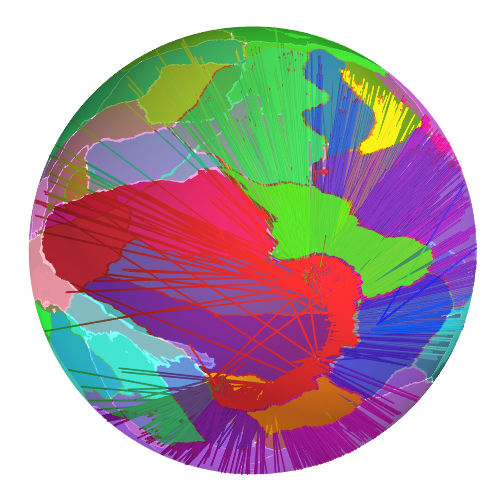}&
\includegraphics[width=0.3\linewidth]{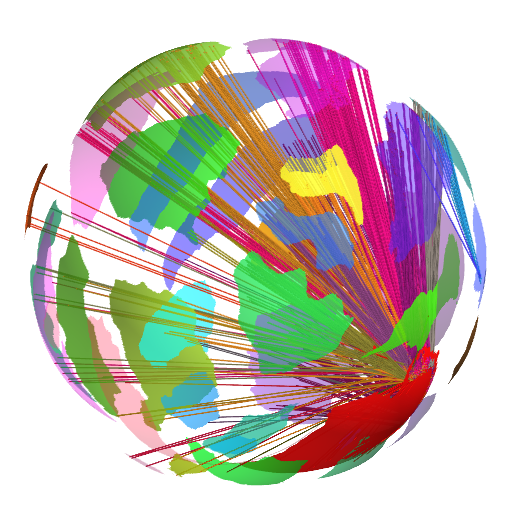}&
\includegraphics[width=0.3\linewidth]{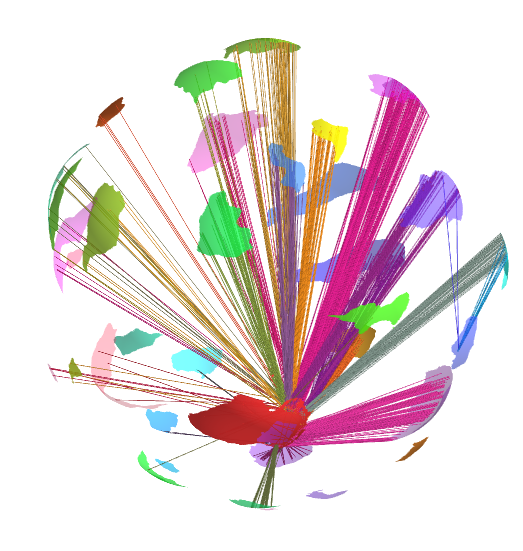}\\
(a) & (b) & (c)\\
\end{tabular}
\end{center}
\caption{\emph{Spherical mapping of the left brain hemisphere with different scales of separation with regions connected by diffusion MRI tractography data}. (a) The spherical mapping of the left brain hemisphere and tracts. (b) The 2x scaled separation of (a). (c) The 4x scaled separation of (a). \emph{Note: Individual tracts are colored with the average RGB values of the color of the pairwise gyri the tracts are connected to. This will be the convention we will use in the spherical separation diagrams to disambiguate tracts.}}
\label{fig:multi_mod_stretced_tracts_sphere}
\end{figure}

\subsection{Structural and Functional Connectivity}
From our pipeline described in Section \ref{sec:pipeline}, we can retrieve structural data (gyri and folds), deterministic tractography data, resting-state functional MRI data, and task-based functional MRI data. These datasets are already registered to the cortical surfaces retrieved from the Mindboggle pipeline and hence, can be explored in the same space, in tandem.

\begin{figure}
\begin{center}
\begin{tabular}{ccc}
\includegraphics[width=0.32\linewidth]{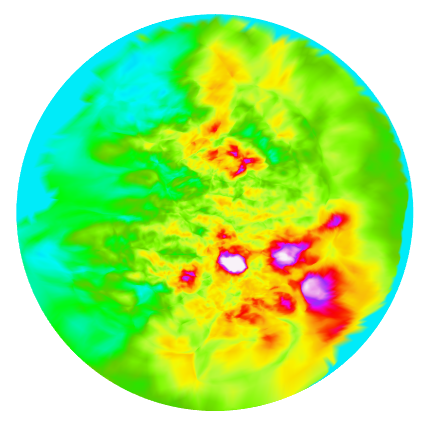}&
\includegraphics[width=0.32\linewidth]{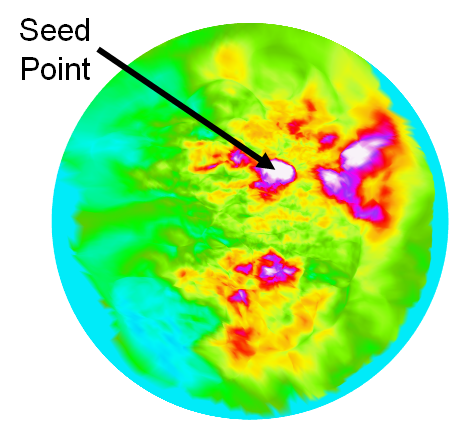}&
\includegraphics[width=0.12\linewidth]{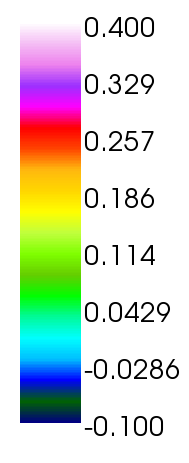}\\
(a) & (b) & \\
\includegraphics[width=0.32\linewidth]{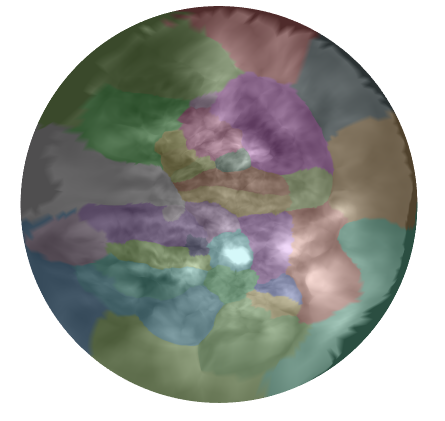}&
\includegraphics[width=0.32\linewidth]{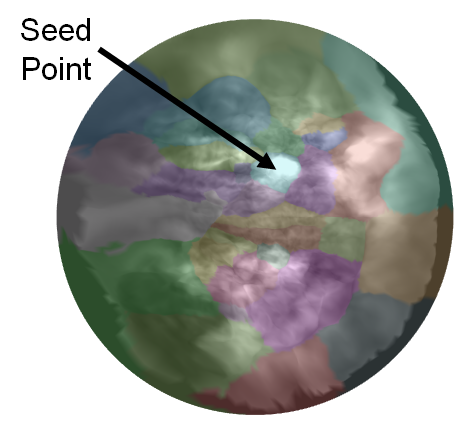}&
\includegraphics[width=0.12\linewidth]{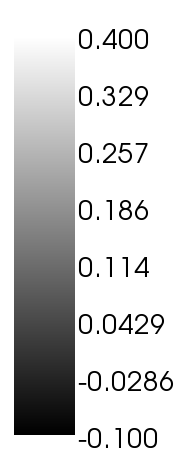}\\
(c) & (d) & \\
\end{tabular}
\end{center}
\caption{\emph{Full correlation functional connectivity map}. (a) and (b) represent the
left and right hemispheres, respectively, with full correlation functional connectivity map, using an exemplar
seed in the right hemisphere (black arrow). (c) and (d) show the gyri labels and the information from (a) and (b) overlaid with a different colormap.}
\label{fig:functional_connectivity}
\end{figure}

\begin{figure}
\centering
\includegraphics[width=1\linewidth]{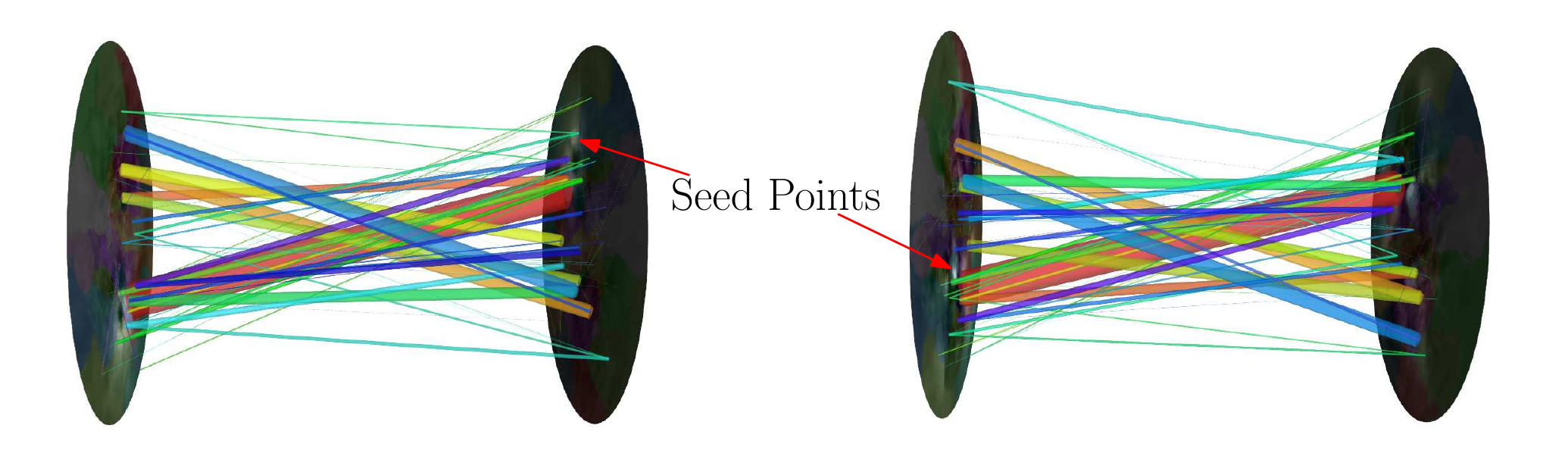}
\caption{\emph{Combined functional connectivity mean gray timecourse and anatomical connectivity coalesced tubes representation}. The gyri information is overlaid with functional connectivity mean gray timecourse representation with two different seed points, along with the coalesced tubes representation of anatomical connectivity from the diffusion MRI tractography data. White regions specify a stronger functional connectivity with the seed point region, as opposed to darker regions which indicate no or weaker connectivity.}
\label{fig:functional_anatomical_tubes}
\end{figure}

We have opted for deterministic tractography, as opposed to probabilistic tractography, in our current work, since it provides the most precise connections between respective regions of interest (ROIs). To compute the structural/anatomical connectivity visualizations in this paper, we have used the corpus callosum as a seed region.

Moreover, in the disk and spherical parametric space, we coalesce tracts into individual coalesced fiber bundles based on the quickbundle fiber clustering algorithm \cite{garyfallidis:2012}; where there are multiple strategies for coalescing tracts, while in this work we coalesce at the endpoints of the cluster. The quickbundle cluster computation is shown in Figure \ref{fig:qb_clustering} and the coalescing of tubes with 2D disks is shown in Figure \ref{fig:anatomical_connectivity}. The network graph corresponding to the anatomical connectivity diagram (in Figure \ref{fig:anatomical_connectivity}) is given in Figure \ref{fig:complete_brain_conn_network} with appropriate region labels.

\begin{figure}
\begin{center}
\begin{tabular}{cc}
\includegraphics[width=0.3\linewidth]{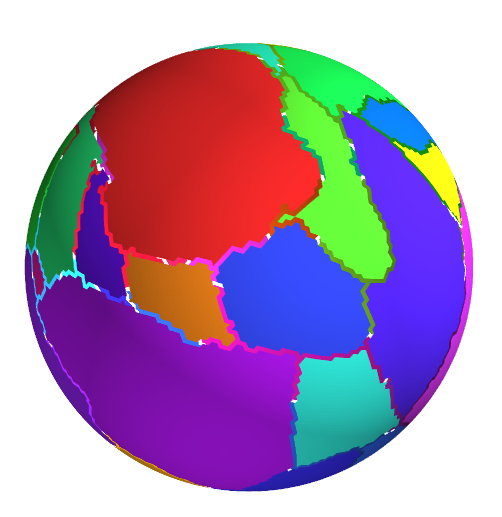}&
\includegraphics[width=0.3\linewidth]{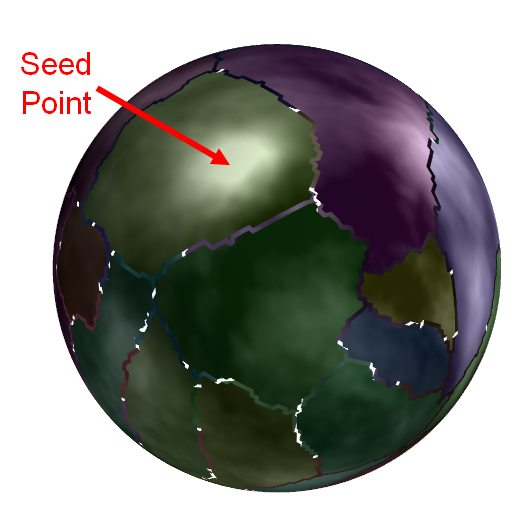}\\
(a) & (b)\\
\includegraphics[width=0.33\linewidth]{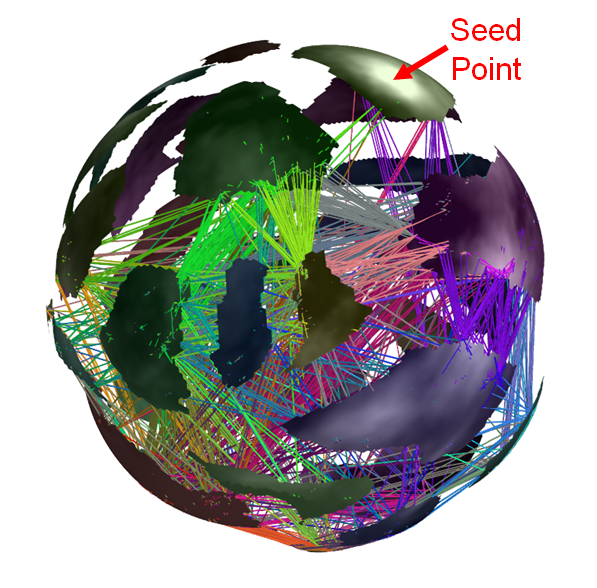}&
\includegraphics[width=0.33\linewidth]{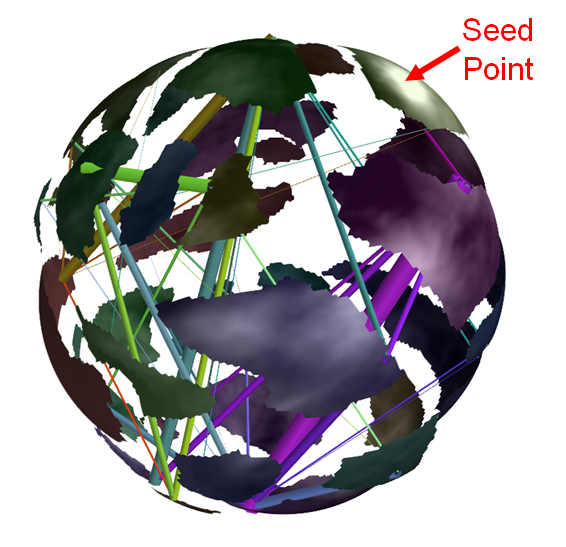}\\
(c) & (d)\\
\end{tabular}
\end{center}
\caption{\emph{Combined spherical mapping of functional connectivity mean gray timecourse and anatomical connectivity}. (a) The functional connectivity information is overlaid onto the anatomical connectivity information from (b) in Figure \ref{fig:multi_mod_stretced_tracts_sphere}. (b) The pairwise gyri area selected from (a) to study the function and anatomical connectivity at a finer granularity.}
\label{fig:functional_anatomical_sphere}
\end{figure}

Using the spherical mapping described in Figure \ref{fig:spherical_map}, we can now map the stretched tracts in different exploded views to delineate the connections clearly, as demonstrated in Figure \ref{fig:multi_mod_stretced_tracts_sphere}. This allows the user to select and interact with individual coalesced fiber bundles in the transformed space.

Furthermore, we can overlay statistical activity maps from functional MRI data onto the cortical surfaces and based on the seed points selected be able to see corresponding high activity area on the cortical surface, as shown in Figure \ref{fig:functional_connectivity}. This can be used to create connectivity diagram in Figure \ref{fig:anatomical_connectivity}, but with functional statistical activity maps overlaid on the two disks, as shown in Figure \ref{fig:functional_anatomical_tubes}. Finally, we can consolidate all these modalities into a complete brain spherical map visualization, shown in Figure \ref{fig:functional_anatomical_sphere}.

\section{CONCLUSION}
We introduced multimodal visualizations of the brain using the disk and spherical area-angle preserving mapping. The rest of the data is presented in different views on this domain. In this work, we have focused on the cortical surface regions. In the future, we will include subcortical regions in our visualizations. We have used deterministic tractography because it provides a better granularity to study the brain than probabilistic tractography.

We are currently working to include subcortical structures and probabilistic tractography with our current visualizations. In the future, we will incorporate information from other modalities, for example, positron emission tomography, magnetoencephalography, and electoencephalography. We will also extend our multimodal visualizations to other organs, such as prostate and pancreas.

\section{ACKNOWLEDGEMENTS}
This project is partially funded by Marcus Foundation Inc. Data were provided by the Human Connectome Project, WU-Minn Consortium (PIs: David Van Essen and Kamil Ugurbil) funded by the 16 NIH Institutes and Centers that support the NIH Blueprint for Neuroscience Research; and by the McDonnell Center for Systems Neuroscience at Washington University.


\end{document}